\documentclass[]{pasj01}

\Received{$\langle$reception date$\rangle$}
\Accepted{$\langle$acception date$\rangle$}
\Published{$\langle$publication date$\rangle$}

\usepackage{graphicx}
\usepackage{tabulary}
\usepackage{natbib}
\bibpunct{(}{)}{;}{a}{}{,}

\newcommand{\M}[0]{\mathsf{M}}
\newcommand{\B}[0]{\mathsf{B}}
\newcommand{\F}[0]{\mathcal{F}}
\newcommand{\I}[0]{\mathsf{I}}
\newcommand{\T}[0]{\mathsf{T}}
\newcommand{\R}[0]{\mathsf{R}}
\newcommand{\A}[0]{\mathcal{A}}
\newcommand{\BB}[0]{\mathcal{B}}

\newcommand{\argmin}{\mathop{\rm arg~min}\limits}

\begin{document}

\title{A Method for Unmasking Incomplete Astronomical Signals: Application to CO Multi-line Imaging of Nearby Galaxies Project}

\author{
Suchetha \textsc{Cooray},\altaffilmark{1,}$^{*}$
Tsutomu T. \textsc{Takeuchi},\altaffilmark{1, 2}
Moe \textsc{Yoda},\altaffilmark{1}
Kazuo \textsc{Sorai}\altaffilmark{3, 4, 5, 6}
}

\altaffiltext{1}{Division of Particle and Astrophysical Science, Nagoya University, Furo-cho, Chikusa-ku, Nagoya 464–8602, Japan}
\altaffiltext{2}{The Research Center for Statistical Machine Learning, The Institute of Statistical Mathematics, 
10-3 Midori-cho, Tachikawa, Tokyo 190-8562, Japan}
\altaffiltext{3}{Department of Physics, Faculty of Science, Hokkaido University, Kita 10 Nishi 8, Kita-ku,
Sapporo 060-0810, Japan}
\altaffiltext{4}{Department of Cosmosciences, Graduate School of Science, Hokkaido University, Kita 10 Nishi 8, Kita-ku, Sapporo 060-0810, Japan}
\altaffiltext{5}{Division of Physics, Faculty of Pure and Applied Sciences, University of Tsukuba, 1-1-1
Tennodai, Tsukuba, Ibaraki 305-8571, Japan}
\altaffiltext{6}{Tomonaga Center for the History of the Universe (TCHoU), University of Tsukuba, Tsukuba, Ibaraki 305-8571, Japan}

\email{cooray@nagoya-u.jp}


\maketitle

\begin{abstract}

    Photometric surveys have provided incredible amounts of astronomical information in the form of images. However, astronomical images often contain artifacts that can critically hinder scientific analysis by misrepresenting intensities or contaminating catalogs as artificial objects. These affected pixels need to be masked and dealt with in any data reduction pipeline. In this paper, we present a flexible, iterative algorithm to recover (unmask) astronomical images where some pixels are lacking. We demonstrate the application of the method on some intensity calibration source images in CO Multi-line Imaging of Nearby Galaxies (COMING) Project conducted using the 45m telescope at Nobeyama Radio Observatory (NRO). The proposed algorithm restored artifacts due to a detector error in the intensity calibration source images. The restored images were used to calibrate 11 out of 147 observed galaxy maps in the survey. The tests show that the algorithm can restore measured intensities at sub 1\% error even for noisy images (SNR = 2.4), despite lacking a significant part of the image. We present the formulation of the reconstruction algorithm, discuss its possibilities and limitations for extensions to other astronomical signals and the results of the COMING application.

\end{abstract}

\section{Introduction} \label{sec:intro}

    In the last decades, we have seen a huge influx of astronomical data in the form of imaging. The Sloan Digital Sky Survey \citep[SDSS;][]{SDSS12_2000AJ....120.1579Y, SDSS3_2011AJ....142...72E, SDSS4_2017AJ....154...28B} spearheaded the various photometric surveys that include the Dark Energy Survey \citep[DES;][]{DES_2016MNRAS.460.1270D}, Hyper Supreme-Cam \citep[HSC;][]{HSC_10.1117/12.926844} and Kilo-Degree Survey \citep[KiDS;][]{KiDS_2013ExA....35...25D}. These surveys have allowed us to have insights on a wide range of topics, from inflation, cosmic expansion to galaxy evolution. In the next decade, with the upcoming projects like the Legacy Survey of Space and Time \citep[LSST;][]{LSST_2019ApJ...873..111I}, we will be in the realm of big data to achieve a multitude of science goals. State-of-the-art CCDs and CMOS detectors will then be essential in obtaining large amounts of information through astronomical imaging.

    However, astronomical imaging is challenging and is uncommon for the CCD/CMOS images to be clean of all errors. Image sensors are a matrix of pixels, and we cannot use every pixel optimally, which leaves us with some unsatisfactory measurements \citep{Janesick2001}. These unexpected artifacts can be a result of broken pixels/columns of pixels, saturation bleeds, or diffraction patterns. In addition to detector artifacts, transients and cosmic rays can further pollute the images. There is a need for these artifacts to be detected and handled. Otherwise, they can contaminate the data and diminish the quality of the information we can get from the surveys.
    
    There are many ways to handle the errors mentioned above. Conservative approaches to deal with the problem would be to discard all the affected pixels during the scientific analysis. Alternatively, we can try to restore the affected pixels. However, the substandard performance of existing restoration methods such as interpolation often prevents the use of restored pixels. The most popular and straightforward way of restoring bad pixels is linear interpolation and is widely implemented in astronomical software such as Image Reduction and Analysis Facility (IRAF) \citep{massey1997user}. 
    
    There have been many extensions to improve the performance of interpolation methods. \cite{Sakurai2001} extended the interpolation technique that conserves photon number counts in astronomical imaging. \cite{Popowicz_2013} introduces a new method known as biharmonic interpolation with relative success and compares some of the existing interpolation methods for bad pixel correction in astronomical images. However, fundamental problems with interpolation exist in all these methods. Interpolation can obtain acceptable results when the missing area is relatively small compared to a large part with information surrounding it. On the contrary, if the missing region is larger than the scale of the reconstructed structures, interpolation fails. Therefore, we suggest that we need to explore fundamentally different and more flexible methods to interpolation for astronomical image reconstruction.
    
    In this paper, we present an extrapolation technique for the reconstruction of masked signals, that are in the form of images. We will analyze the applicability of the algorithm in the context of astronomical imaging and demonstrate a real-world application. Throughout the article, we will use the term signal interchangeably to refer to images as they are a two-dimensional measurement of the signal.
    
    Firstly, let us mathematically define the problem we solve. We suppose that there exists an idealistic astronomical signal for every observed faulty astronomical signal. Then, let the ideal image be $f$, and the observed (masked) be $g$. We can relate the two images/signals mathematically as follows;
    \begin{equation} \label{eq:1}
        g(x, y) = \M_\Gamma(x, y) f(x, y) ,
    \end{equation}
    where $\M_\Gamma$(x, y) is defined as;
    \begin{equation} \label{eq:mask}
        \M_\Gamma (x, y) = \left \{ \begin{array}{ll}
                1  &  \  \textrm{if } (x, y) \in \Gamma \\
                0   &  \ \textrm{elsewhere} .
        \end{array} \right. 
    \end{equation}
    $\M_\Gamma$ is a distortion operator that masks a part of the image. The unmasked regions of the image become the observed/unaffected region $\Gamma$. With any of the existing reconstruction methods, we are solving the following equation,
    \begin{equation} \label{eq:solving_eq}
        f =  {\argmin_{f^{\prime}} } \left\{ \left|\left|\M_\Gamma f^{\prime} - g \right|\right|^2 \right\},
    \end{equation}
    which is an \textit{inverse problem}. The solution is obtained by effectively inverting the masking operator, where what is of interest ($f$) is inferred from the observable ($g$).
    
    To solve the above inverse problem of Eq. (\ref{eq:solving_eq}), one can formulate the maximum-likelihood estimator (MLE) for some assumed noise distribution in the observable. For example, we consider a two-dimensional signal (image) with noise that follow $\mathcal{N}\left(0, \sigma_{xy}^{2}\right)$ at an observed pixel $(x,y)$. Then the log-likelihood $L_{\textrm{Gaussian}}$ to be maximized can be written as,
    \begin{equation}
        L_{\textrm{Gaussian}} = - \sum_{x, y} \frac{1}{2\sigma_{x y}^{2}}(g - \M_\Gamma f)^{2} .
    \end{equation}
    Therefore, the reconstruction of partial signals with Gaussian noise is equivalent to maximizing the above log-likelihood under the MLE technique. Various optimization algorithms can be employed to solve the above problem.
    
    However, many inverse problems are also \textit{ill-posed} problems, and may not have a unique explicit solution as the masking operator can be singular. To obtain a reasonable solution to an ill-posed inverse problem, we need some a priori information about the sought after solution.

    One method for solving the inverse problem is an iterative extrapolation algorithm proposed by A. Papoulis and R. W. Gerchberg. \citet{Papoulis1975} described an iterative algorithm for estimating the entire one-dimensional analytic function from parts of a function in real space under assumptions in the Fourier space. \citet{Gerchberg1974} introduced an algorithm for super-resolution beyond the diffraction limit by the concept of iterative "error energy" reduction. Despite independently presented by the authors, in essence, they are the same and is now often known as the Papoulis-Gerchberg algorithm. 
    
    The two works were based on a fundamental property of the Fourier transform that a function with finite support (domain) in real space will require infinite support in the Fourier space and vice versa \citep{Benedicks1985, Cowling1984, Amrein1977}. A signal with finite support in Fourier space is known as a \textit{bandlimited} signal. Such a signal will never have limited support in real space. With the same argument, partial signals in real space can never have finite support in Fourier space. Therefore, imposing a bandlimited assumption can be used to estimate the lacking real-space regions. We note that a definite bandlimited signal requires an infinite domain in real space to express, and such is impossible due to finite resources. However, the concept is an essential idealization of real-world signals.

    The studies by Papoulis and Gerchberg brought a new interest in the 70s, and many papers have since appeared that address various aspects of the bandlimited signal extrapolation problem. \citet{Youla1978} has shown us a more general geometric view of the iterative reconstruction methods providing a natural look at a larger group of similar algorithms. A comparison of different bandlimited extrapolation algorithms for discrete signals is discussed in \citet{Jain1981}. \citet{Huang1984} compares iterative and non-iterative extrapolation algorithms for noisy signals. \citet{Cenker1991} gives an overview and a comparison of reconstruction algorithms for irregular sampling. The extrapolation algorithm itself is a particular case of gradient descent, which are cases of more general methods of proximal splitting \citep{2009arXiv0912.3522C}. They are just a handful of works on the bandlimited signal extrapolation problem. The study is a well-established area in information sciences and signal processing despite been relatively unknown in the astronomical community.

    In this work, we demonstrate a framework suited for the reconstruction of astronomical images based on the Papoulis-Gerchberg algorithm. We show an application to the distorted intensity calibration source images of the COMING (CO multi-line imaging of nearby galaxies) Project \citep{Sorai2019}. COMING Project is a legacy project done with the 45m radio telescope at Nobeyama Radio Observatory (NRO)\footnote{Nobeyama Radio Observatory is a branch of the National Astronomical Observatory of Japan, National Institutes of Natural Sciences.}, where they mapped 147 nearby galaxies in $^{12}$CO, $^{13}$CO, and the C$^{18}$O lines simultaneously using the intermediate frequency band of the telescope's multi-beam receiver, FOREST \citep{Minamidani_2016}. During some observations of the intensity calibration source, the detector unexpectedly shifted its reference frequency, producing artifacts in the velocity integrated intensity maps. Despite the mechanism for the artifact being different from usual astronomical images, the results are similar to bad pixels in images from CCDs and CMOS sensors. CO multi-line maps of 11 galaxies out of the total observed were considered inoperable due to the issue in the calibration source images. We restored the distorted images using the proposing reconstruction algorithm, and those images calibrated the observed galaxy maps. We will later describe the tests done on the suitability of the algorithm for this application. 

    There are two primary motivations for this paper. Primarily, we present a mathematically consistent reconstruction technique inspired by information sciences for astronomical signals. We analyze the possibilities and the limitations of such algorithms on astronomical/signals. The secondary motivation is to inspire more applications of the reconstruction algorithm as an approach to make most out of the information available. We achieve this by presenting an example application of the reconstruction algorithm on the real-world astronomical images (i.e., COMING).

    The paper will be structured as follows. In Section \ref{sec:recon_algo}, we present the theory, formulation, and discuss the convergence of the reconstruction algorithm. We then examine the erroneous data, conduct tests of the reconstruction algorithm, and describe the restoration of the distorted COMING intensity calibration source images in Section \ref{sec:coming}. Followed by the application in the COMING project is a discussion on the possibilities and limitations of the algorithm in Section \ref{sec:discussion}. Lastly, in Section \ref{sec:conclusion}, we consider the implications of this work and possible applications of the presented method in other areas of astronomical measurements.

\section{Reconstruction Algorithm} \label{sec:recon_algo}
    In this section, we provide the mathematical steps of the reconstruction, proceeded by an explanation of the procedure for possible implementation on the computer. Afterward, we demonstrate a proof for the existence of the expected unique solution (the complete image) and the convergence by the iterative algorithm to it. Table \ref{table:symbols} summarizes the symbols used throughout this paper.
    
    \begin{table}
        \tbl{List of symbols}{
        \centering
        \begin{tabulary}{\linewidth}{l L} 
            \hline
            Symbol & Description\\
            \hline \hline
            $f$ & Ideal complete signal \\
            $g$ & Observed incomplete signal \\
            $\Gamma$ & Observed region of the signal \\
            $\M_\Gamma$ & Distortion operator which masks the signal in the region outside of $\Gamma$\\
            $\F$ & Fourier transform \\
            $\F^{-1}$ & Inverse Fourier transform \\
            $F$ & Fourier transform of $f$ \\
            $G$ & Fourier transform of $g$ \\
            $\B$ & Bandlimiting operator: $\F^{-1} \mathsf{\beta} \F$ \\
            $\Omega$ & Domain of the Fourier space for the bandlimit \\
            $\mathsf{\beta}_{\Omega}$ & Band selecting operator which filters Fourier coefficients outside $\Omega$ \\
            $\tilde{f}$ & Bandlimited signal of $f$ \\
            $\tilde{F}$ & Filtered Fourier coefficients of $f$ \\
            $f_n$ & $n^{\textrm{th}}$ estimation of the complete signal \\
            $g_n$ &  Combination of mask region in $f_n(x, y)$ and observed $g$ \\
            $F_n$ & $n^{\textrm{th}}$ extrapolation of $F$ \\
            $G_n$ & Fourier transform of $g_n$ \\
            $\T$ & Iterative operator: $g + (\I - \M_{\Gamma}) \B$ \\
            $\R$ & $(\I - \M_{\Gamma}) \B$ \\
            $r_n$ & residual at the $n^{\textrm{th}}$ iteration: $g_n - f$ \\
            $I$ & Measured intensity \\
            $e$ & Dimensionless intensity reconstruction error \\
            \hline
        \end{tabulary}}
        \label{table:symbols}
    \end{table}
    
    \subsection{Theory}
         Let us begin by defining the discrete Fourier transform (FT) of $f(x, y)$ for a $M \times N$ image;
        \begin{equation}
            F(u, v) = \F f(x, y) \equiv \sum ^{M-1}_{x=0} \sum ^{N-1}_{y=0} f(x, y) e^{-2 \pi i (\frac{ux}{M} + \frac{vy}{N})},
        \end{equation} 
        and its inverse would be;
        \begin{equation}
            f(x, y) = \F^{-1} F(u, v) \equiv \frac{1}{NM} \sum ^{M-1}_{u=0} \sum ^{N-1}_{v=0} F(u, v) e^{2 \pi i (\frac{ux}{M} + \frac{vy}{N})}.
        \end{equation}

        Reconstruction of $f$ (complete signal) from $g$ (incomplete signal) is done by imposing some assumed knowledge of $f$. Then let $\B$ be an operator that contains the knowledge of $f$. When $\B$ operates on a signal, it imposes the constraints on the signal. Mathematically we can write the above as,
        \begin{equation}
            \tilde{f}(x, y) = \B f(x, y),
        \end{equation}
        where $\tilde{f}$ is the signal obeying the constraints in $\B$. When $\tilde{f} = f$, $\B$ is just the identity. Therefore, $\tilde{f} = f$ implies that $f$ is a signal that obeys the constraint information encapsulated in $\B$. The constraint of $f$ for the reconstruction algorithm is that the signal is bandlimited. A bandlimited signal is a square-integrable function whose Fourier transform is zero outside a bounded interval. That is to say that the signal can be fully expressed in Fourier space by a finite domain (finite support). Bandlimitedness is a very natural property for most real signals as there are no indefinitely high-frequency components in real-world signals. For the case where $f$ is a bandlimited signal, $\B$ is a bandlimiting operator defined as $\B = \F^{-1} \mathsf{\beta} \F$ where $\mathsf{\beta}$ is a band selecting operator that lets pass only certain frequencies. We can express the ideal bandlimiting operator as the following;
        \begin{equation}
            \tilde{F}(u, v) = \mathsf{\beta} (u, v) F(u, v),
        \end{equation}
        where $F$ is the Fourier transform of $f$ and $\tilde{F}$ are the filtered Fourier coefficients. We can define the frequencies that is allowed to pass through as $\Omega$. The above can then be used to define $\mathsf{\beta}_{\Omega}$ as,
        \begin{equation} \label{eq:bandlimit_1}
            \mathsf{\beta}_{\Omega}(u, v) = \left \{ \begin{array}{ll}
                    1  &  \quad \textrm{if } (u, v) \in \Omega \\
                    0  &  \quad \textrm{otherwise           .}
            \end{array} \right.
        \end{equation}
        
        In the context of images, let us define a 2D ideal low-pass filter. For some positive finite values $U$ and $V$ less than $M$ and $N$ respectively, we can define $\Omega$ as,
        \begin{equation} \label{eq:bandlimit_2}
            \Omega = \{ (u, v) \ | \ ( |u| \leq U \ \& \ |v| \leq V ) \}.
        \end{equation}
        For an image (discrete signal) to be reconstructed, $U$ and $V$ must be less than $M$ and $N$, respectively. The reason is that an image is by design bandlimited due to the finite sum in the discrete Fourier transform. In the context of images, the signal needs to be bandlimited with a smaller Fourier domain than the domain defined by the size of the image. More will be discussed on this aspect in Section \ref{sec:discussion}. However, the algorithm is not limited to a 2D ideal low-pass filter, and therefore, the bandlimiting operator $\B$ can be defined differently according to the application.
        
        The iterative extrapolation algorithm begins by estimating the $0^{\textrm{th}}$ Fourier transform of $g$,
        \begin{equation} \label{eq:first_estimation_1}
            G_0 (u, v) = G(u, v) \equiv \F \ [\M_\Gamma (x, y) g(x, y)].
        \end{equation}
        The first extrapolation $F_{1}(u, v)$ is given by;
        \begin{equation} \label{eq:first_estimation_2}
            F_1 (u, v) = \beta_{\Omega}(u, v) G_0 (u, v).
        \end{equation}
        Then the first estimated signal/image will then be the inverse Fourier transform of $F_1 (u, v)$, 
        \begin{equation}
            f_1 (x, y) = \F^{-1} \  F_1(u, v).
        \end{equation}
        As the algorithm is iterative by nature, we use the first estimation $f_1(x, y)$ for the next estimation. We replace the segment of  $f_1(x, y)$ in the region $\Gamma$ with the observed $g(x, y)$.
        \begin{eqnarray}  \label{eq:combine_1}
            g_1(x, y) &=& \M_{\Gamma}g(x, y) + [\I - \M_{\Gamma}] f_1(x, y) \nonumber \\
                              &=& f_1(x, y) + \M_{\Gamma} [g(x, y)-f_1(x, y)],
        \end{eqnarray}
        where $\I$ is just the identity mapping. The derived equation above is essentially,
        \begin{equation} \label{eq:combine_2}
            g_1(x, y) = \left \{ \begin{array}{ll}
                    g(x, y)  &  \quad \textrm{if } (x, y)  \in \Gamma \\
                    f_1(x, y)   &  \quad \textrm{elsewhere    .}
            \end{array} \right.
        \end{equation}
        The first iteration ends by the final step of finding $G_1(u, v)$;
        \begin{equation} 
            G_1(u, v)= \F \  g_1(x, y).
        \end{equation}
        The above explained process is repeated. We show the procedure for the \textit{n}$^\textrm{th}$ iteration as follows;

\begin{figure*}
            \centering
            \includegraphics[width = 0.7\textwidth] {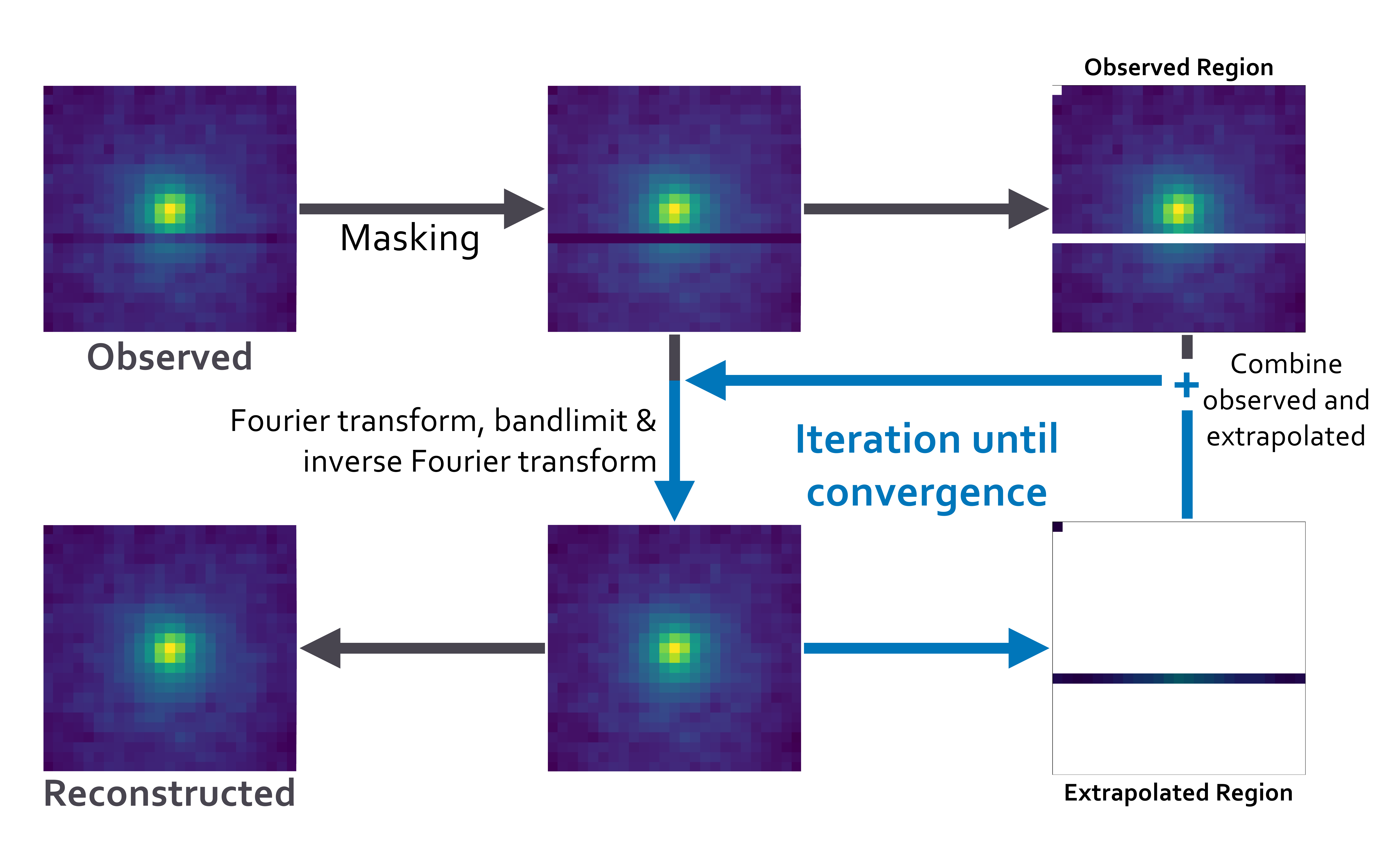}
            \caption{{The schematic diagram of the procedure for the reconstruction algorithm is shown. The process begins with the observation with the faulty band of pixels. Then the defective pixels are masked. FT is done using the pixels with information.  The FT of the image passes through an ideal band-pass filter based on $\Omega$ after which, is transformed back to real space. The extrapolated region and the initially observed information is combined. Then the combined image is used for FT from which the process is repeated until a condition to stop the iteration is satisfied. Once the iteration ends, we have the reconstructed image.}}
            \label{fig:algo_schematic_diagram}
        \end{figure*}

        The \textit{n}$^\textrm{th}$ extrapolation in Fourier space is estimated by, 
        \begin{equation} \label{eq:bandlimiting}
            F_n (u, v) = \beta_{\Omega}(u, v) G_{n-1} (u, v),
        \end{equation}
        then we inverse Fourier transform,
        \begin{equation} \label{eq:ifourier}
            f_n (x, y) = \F^{-1} \  F_n(u, v).
        \end{equation}
        The unmasked segment in $f_n(x, y)$ is replaced with $g(x, y)$ to obtain the \textit{n}$^\textrm{th}$ estimation of the signal as,
        \begin{equation} \label{eq:combine}
            g_n(x, y)  = \M_{\Gamma}g(x, y) + [\I - \M_{\Gamma}] f_n(x, y).
        \end{equation}
        If the iteration does not terminate, we Fourier transform $g_n(x, y)$ as to begin the next extrapolation,
        \begin{equation} \label{eq:fourier}
            G_n(u, v)= \F \  g_n(x, y).
        \end{equation}
        We can then show the above iterative extrapolation procedure for the \textit{n}$^\textrm{th}$ iteration as,
        \begin{equation}
            g_n(x, y) = g(x, y) + [\I - \M_{\Gamma}] \B g_{n-1} = \T g_{n-1} (x, y) .
        \end{equation}
        where 
        \begin{equation} \label{eq:T_operator}
            \T = g + [\I - \M_{\Gamma}] \B ,
        \end{equation}
        is an iterative operator. As $n$ tends to infinity, the estimation and the solution will converge and we will obtain the original signal. That is,
        \begin{equation} \label{eq:convergence}
            g_n(x, y)\rightarrow f(x, y) \textrm{ as } n \rightarrow \infty .
        \end{equation}
        For increasing $n$, we expect the operator $\T$ to converge to zero. We will discuss in Section \ref{sec:convergence} how the iterative extrapolation operator is convergent to zero as the number of iteration tends to infinity. In implementations, a condition to end the iteration is when $\T$ is small enough. In other words, when the difference between the successive estimations is infinitesimal. Figure \ref{fig:algo_schematic_diagram} shows the schematic diagram in a flowchart format for an easy understanding of the algorithm.
        
         \label{sec:theory}

    \subsection{Convergence of the Algorithm} \label{sec:convergence}
        We have shown above an algorithm that reconstructs $f$ from $g$ under some known information about $f$. Shown below is a discussion of this convergence.
        
        Let us first introduce the notion of a \textit{nonexpansive} operator. An injective operator $\A :X \rightarrow X$ is nonexpansive if
        \begin{equation}
            \| \A x - \A y \| \leq \| x-y \| \textrm{ for } x,y \in X .
        \end{equation}
        Furthermore, the operator $ \A $ is \textit{strictly nonexpansive} if the above equity holds only for $x=y$ or similarly,
        \begin{equation}
            \| \A x- \A y \| < \| x-y \| \textrm{ for } x,y \in X .
        \end{equation}
        Now, if an operator $\A$ is nonexpansive (strictly nonexpansive), then for an arbitrary $x \in X$, $\| \A x\| \leq \| x \|$ ($\| \A x \| < \| x \|$). 
        It is possible to relate the spectral radius $\rho ( \A )$ and the norms as, $\| \A \| \geq \rho (\A)$. The proof for the relation is described in the Appendix \ref{sec:appendix_nonexpansive}. We then have that the spectral radius $\rho (\A)$ for $\A$ nonexpansive (strictly nonexpansive) should be $\rho (\A) \leq 1$ ($\rho (\A)<1$).
        
        As seen from above, for a strictly nonexpansive operator, the spectral norm should be less than unity. In this sense, a strictly nonexpansive operator has a "gain" of less than unity. Thus, it is reasonable for a strictly nonexpansive operator to be also called a \textit{contraction mapping}.
        
        If an operator is a contraction mapping, it guarantees the uniqueness and the existence of a \textit{fixed point} \citep{Smart1980, Schafer1981}. The \textit{fixed point} of an operator $\A:X \rightarrow X$ is $\A x=x$ for $x \in X$. The above statement is called the \textit{contraction mapping theorem} or \textit{fixed-point theorem} and is a fundamental result of functional analysis. In our definition of $\A$, we supposed that $\A$ is injective. For an injective operator $\A :X \rightarrow X$, $\A x = \A y$ implies $x = y$ for $x,y \in X$ and above argument remains valid, ensuring the uniqueness and existence of the fixed point \citep[e.g.,][]{Smart1980}.
        
        If we can show that the iterative extrapolation operator $\T$ from the previous section is strictly nonexpansive ($\rho (\T) <1$), then $\T$ is convergent. We can thus prove that the iterative algorithm discussed in this paper will have a unique solution $f$ for $\T f=f$ that is given by,
        \begin{equation}
            f = \lim_{n \rightarrow \infty} \T ^{(n)} g ,
        \end{equation}
        for $g$ in the same image space as $f$. The statement is similar to the convergence shown in equation (\ref{eq:convergence}). We mathematically show how the operator $\T$ is strictly nonexpansive, a contraction or has a fixed point in Appendix \ref{sec:appendix_convergence}.
        
        To ensure a unique fixed point for $\T$ in the context of signal restoration, we need to consider the size of the masked region and the number of Fourier components to be reconstructed. Considering the real and imaginary parts as two measurements, let the number of independent values allowed through $\M_{\Gamma}$ be $L$ and for $\beta_{\Omega}$ be $K$.\footnote{We remind the reader that for a real signal with $N$ measurements, $N/2-1$ complex Fourier coefficients are redundant due to the conjugate symmetry property of the FT. For a complex image, the number of measurements is doubled (real and imaginary), and the redundancy in the Fourier coefficients breaks down. Even in this case, inequality should still read $L\geq K$ in our definition.} To determine a unique solution, we need to estimate $K$ nonzero coefficients from $L$. In other words, the convergence to a unique solution requires the condition, $L \geq K$. For example, a real image with a single unmasked pixel ($L=1$) can guarantee a unique solution only if we assume an image of constant values ($K=1$).
        
        Let us now consider the residual at each iteration $i$ as $r_i=g_i-f$, where $g_0 = g$ and $r_0$ would be the initial residual. We could then write,
        \begin{equation}
            r_i = \T r_{i-1} .
        \end{equation}
        For $i$ iterations, we have
        \begin{equation}
            r_i = \T^{(i)} r_0 .
        \end{equation}
        Following the previous argument, we can now talk about the norm of the residual for the i$^\textrm{th}$ iteration,
        \begin{equation}
            \| r_i \| \leq \| \T^{(i)} \| \cdot \| r_0 \| .
        \end{equation}
        As we have shown that $\T$ is a contraction, the norm of the residual at the i$^\textrm{th}$ iteration will decrease with each iteration. It is then clear that $\T^{(i)}$ tends to zero monotonically with the norm of the residual. We then have, by definition, that $\T$ is convergent \citep{Varga2000}. This error-reducing property is what we look for in an iterative algorithm, and \cite{Ferreira1994a} theoretically discusses more on the upper and lower bounds of the above residual.
        
        We note that the above-explained algorithm can be strictly applicable only to {\sl noiseless} signals, and under the existence of noise, the converged solution may not be the maximum-likelihood solution. 
        However, \citet{Sanz1983} have shown analytically that the procedure can produce good approximations for moderately noisy signals. 
        We extensively test actual reconstruction performance with noisy synthetic simulations in Section \ref{sec:noise_tests}.

\section{Application in COMING Project} \label{sec:coming}

    We demonstrate an application of the reconstruction algorithm (explained above) to the distorted images in the COMING project. We will first describe the faults in the calibration source images and their causes (Section \ref{sec:data}). We then discuss the tests with generated distorted mock images. The mock images based on complete observations are masked and then restored to understand the performance of the algorithm (Section \ref{sec:noise_tests}). Afterward, we test our algorithm by reconstruction of artificially masked complete observed images (Section \ref{sec:error_free_reconstruction}). Lastly in Section \ref{sec:reconstruction}, we reconstruct the incomplete COMING calibration source images.

\subsection{Faulty Images of COMING Calibration Sources} \label{sec:data}
    This section describes the incomplete image data reconstructed using the algorithm explained in Section \ref{sec:recon_algo}. The images are of the intensity calibration sources of the COMING project, which are observations of the standard intensity object IRC+10216 (($\alpha$, $\delta$)$_{\textrm{B1950.0}}$ = (09$^h$45$^m$15$^s$.0, +13$\degree$30'45". 0)). Observations were done at rest-frame frequencies of 110.201353 GHz and 115.271202 GHz for $^{12}$CO and $^{13}$CO, respectively. The FOREST detector on the 45m telescope at NRO has four beams with two polarization each. Each polarization was observed in both CO lines ($^{12}$CO and $^{13}$CO), and thus, for every standard object observation, 16 data cubes were obtained. The observations were then integrated across the frequencies to obtain intensity maps of size 25 $\times$ 25 pixels. All of the reconstructed images shown in this paper are of the same celestial object, and so the center of the images correspond to (RA, DEC) = (146.3125\degree, 13.5125\degree), and each pixel corresponds to $2.083\times10^{-3}$ \degree.

\begin{figure}
        \centering
        \includegraphics[width=\linewidth]{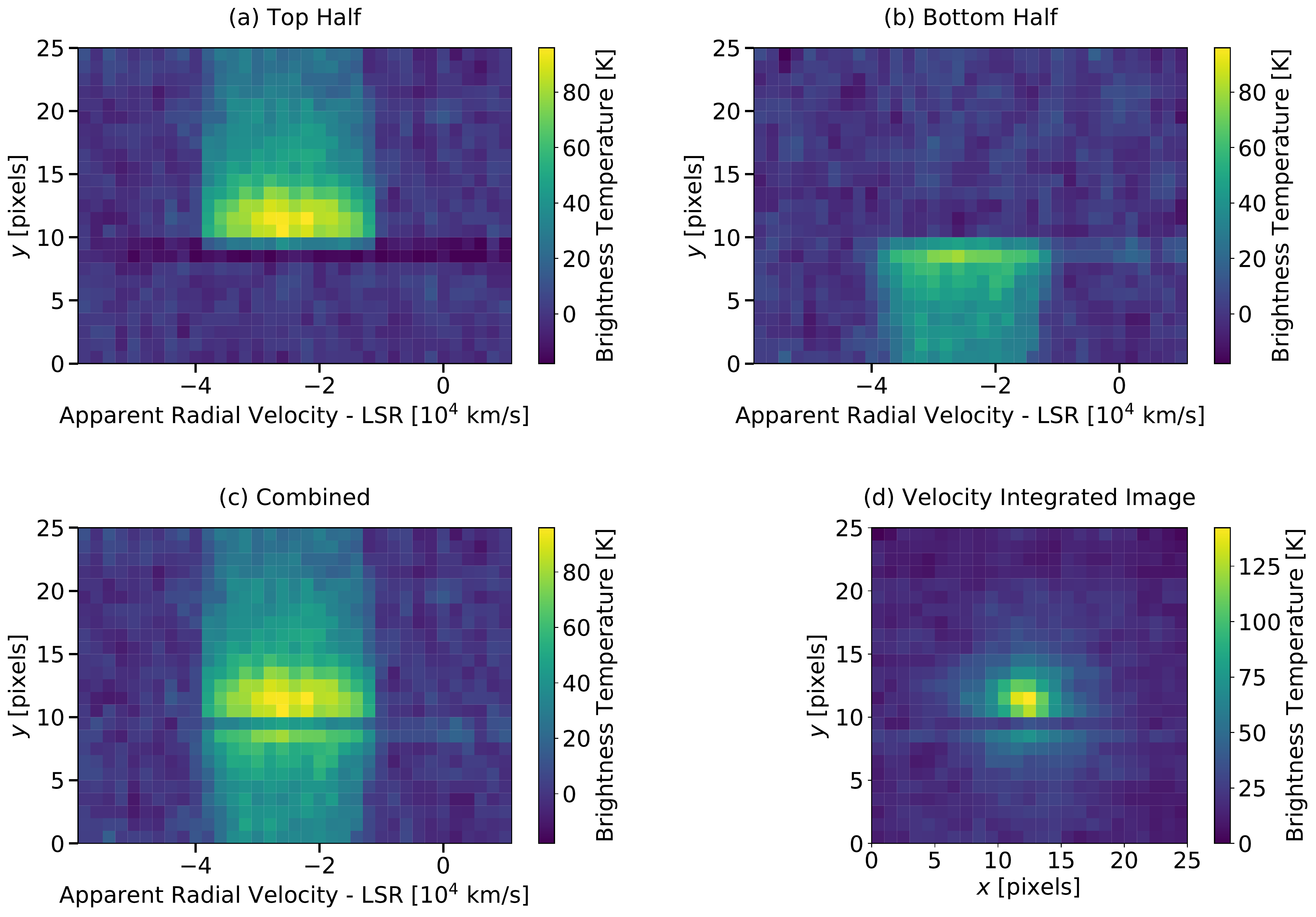}
        \caption{(a) is the top half of the beam observed. (b) is the bottom half and (c) is the combined of (a) and (b) by correcting the frequency shift. (d) is the combined integrated frequency map.}
        \label{fig:frequency_shift}
    \end{figure}

    On six sets of observations done on the standard intensity source from March to April 2018, the reference signal shifted during the on-the-fly scans of the FOREST detector. The shift resulted in part of the signal separated in the frequency (apparent radial velocity) direction of the observed 3-dimensional block. This separation was then manually corrected based on the intensity peak and shape. However, the bordering row of pixels showed unexpected artifacts as a result of the shift correction. Figure \ref{fig:frequency_shift} shows an example of the shifted and the combined block, together with the frequency integrated image. The band of pixels with seemingly artificial intensity values prevented the use of these observations for calibration of galaxy maps. We consider the reconstruction of these affected images using the algorithm described in Section \ref{sec:recon_algo}. The defective pixels in the calibration source images are masked, and the reconstruction algorithm applied.
    
    In addition to this error, out of the 16 arrays of each observation, the observations with beams 2 and 3 (arrays 3 to 6 of $^{12}$CO and $^{13}$CO observations) had a significant part of the image without information. However, this error does not affect the intensity calibration significantly as the lacking region was outside the main signal region. We reconstruct this region together with the region with the artifact. An example for set of 16 calibration source images in both $^{12}$CO and $^{13}$CO observations is shown on the left side of Figure \ref{fig:orig_and_recon}.

\subsection{Reconstruction Tests with Simulated Noisy Observation} \label{sec:noise_tests}

    We conduct tests with simulated images to analyze the performance of the reconstruction algorithm under noise. As discussed in Section \ref{sec:convergence}, noise affects the reconstruction performance because when noise is introduced to a pure bandlimited signal, the signal deviates from the assumptions, diminishing the performance. 
    
    We generate mock image data with varying levels of noise to test the reconstruction algorithm. The model signal for the mock images is the best fit function for the highest SNR complete observation described. We fitted 2-dimensional Gaussian, Airy \citep{Airy_1835}, and Moffat \citep{Moffat_1969} functions to the observed image. Moffat function produced the best fit, according to the information criterion. We thus assume the fitted Moffat function with additive random noise as mock images. 
    
    Noise value for each pixel is picked from a Gaussian distribution with a fixed standard deviation according to the signal-to-noise ratio \citep[SNR: e.g.,][]{birney2006observational}. The SNR was defined to be; 
    \begin{equation}
        \textrm{SNR} = \frac{N_s}{\sqrt{N_s + N}},
    \end{equation}
    where $N_s$ is the $\ell_1$-norm of the noiseless signal and $N$ is the $\ell_1$-norm of the noise from various factors that include the atmospheric conditions and read noise.
    
    A realistic mask, according to the COMING calibration source images, is applied to the generated images. For the distorted images, one or two rows of pixels at row numbers 8, 9, 16, 17 were masked. We found that artifacts in the $16^{\textrm{th}}$ / $17^{\textrm{th}}$ pixel rows affects the calibration less compared to the artifacts in $8^{\textrm{th}}$ / $9^{\textrm{th}}$ rows because the region is further away from the signal peak. As a reconstruction test, we consider the most difficult situation for the COMING calibration data. Therefore, the $9^{\textrm{th}}$ and $10^{\textrm{th}}$ rows were masked as simulated faulty observations.
    
    \begin{figure}
        \centering
        \includegraphics[width=0.5\textwidth]{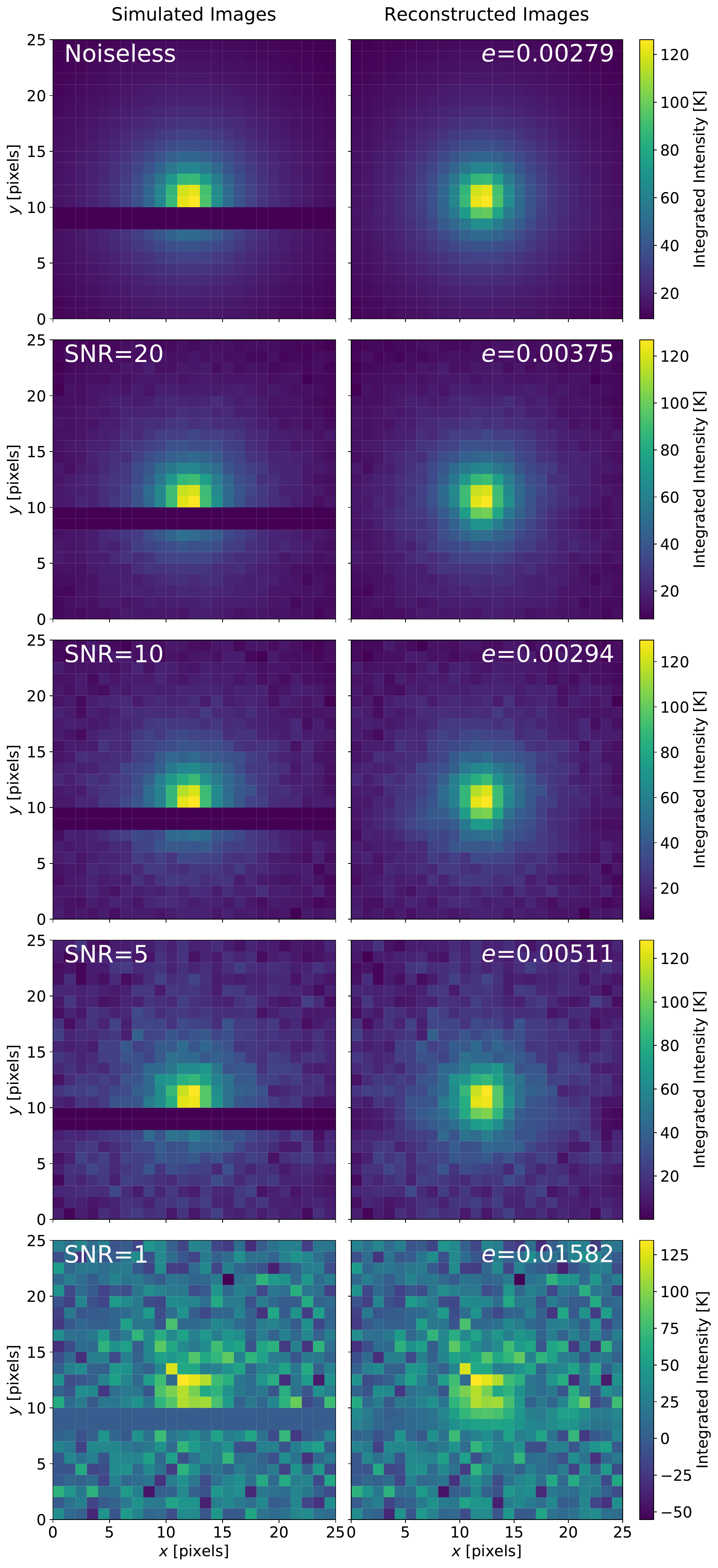}
        \caption{Panels on the left show simulated images with SNR = 1, 5, 10, 20 and noiseless. Panels on the right show their reconstructed versions with the reconstruction error calculated using Eq. (\ref{eq:recon_error}). Reconstruction error values agree with Fig. \ref{fig:mock_reconstruction_error}. Successful reconstruction of the masked region is seen even for the noisiest image (SNR=1).}
        \label{fig:mock_reconstruction_example}
    \end{figure}
        
    We generate 1000 masked images for each SNR and reconstruct them. Figure \ref{fig:mock_reconstruction_example} shows some examples of simulated images with various noise levels and their reconstructions. The bandlimit constraint parameters are determined as described in Appendix \ref{sec:appendix_bandlimit}. The termination criterion for the iteration is when the solution converges, and the normalized difference between the measured intensities of successive estimations is 0.0001. For these images, the convergent criterion was met for $n < 1000$. Such is not computationally heavy on any modern computer with the use of fast Fourier transforms.
    
    The measured intensity is considered to be the $\ell_1$-norm of the images within a $l \times l$ block centered at the peak of the signal. Let $A(x, y)$ for $(x,y = 1,...,25)$ be the pixel value of the image at the location $(x, y)$. When the signal peak is ($x_c, y_c$), the measured intensity $I$ of image $A$ within a block of side $l$ is;
    \begin{equation}
        I(A, l) = \sum_{x=x_c-l}^{x_c+l} \sum_{y=y_c-l}^{y_c+l} A(x,y) .
    \end{equation}
    $l=11$ was adopted for intensity measurement (i.e. $11 \times 11$ block) as described in \citet{Sorai2019}. 
    
    In order to asses the reconstruction performance, we define the dimensionless intensity reconstruction error $e$ as,
    \begin{equation} \label{eq:recon_error}
        e = \sqrt{\frac{|I(\textrm{reconstructed}) - I(\textrm{original})|^{2}}{\sum|I(\textrm{original})|^{2}}} .
    \end{equation}
    Such a metric may not consider pixel-to-pixel reconstruction accuracy such as with normalised root mean square error \citep{Fienup:97}. However, here we introduce the above metric because this is standard for the particular case of flux calibration. In this case, the contribution of the noise to the total amplitude should be considered in the reconstructed intensities.
    
\begin{figure}
        \centering
        \includegraphics[width=0.95\linewidth]{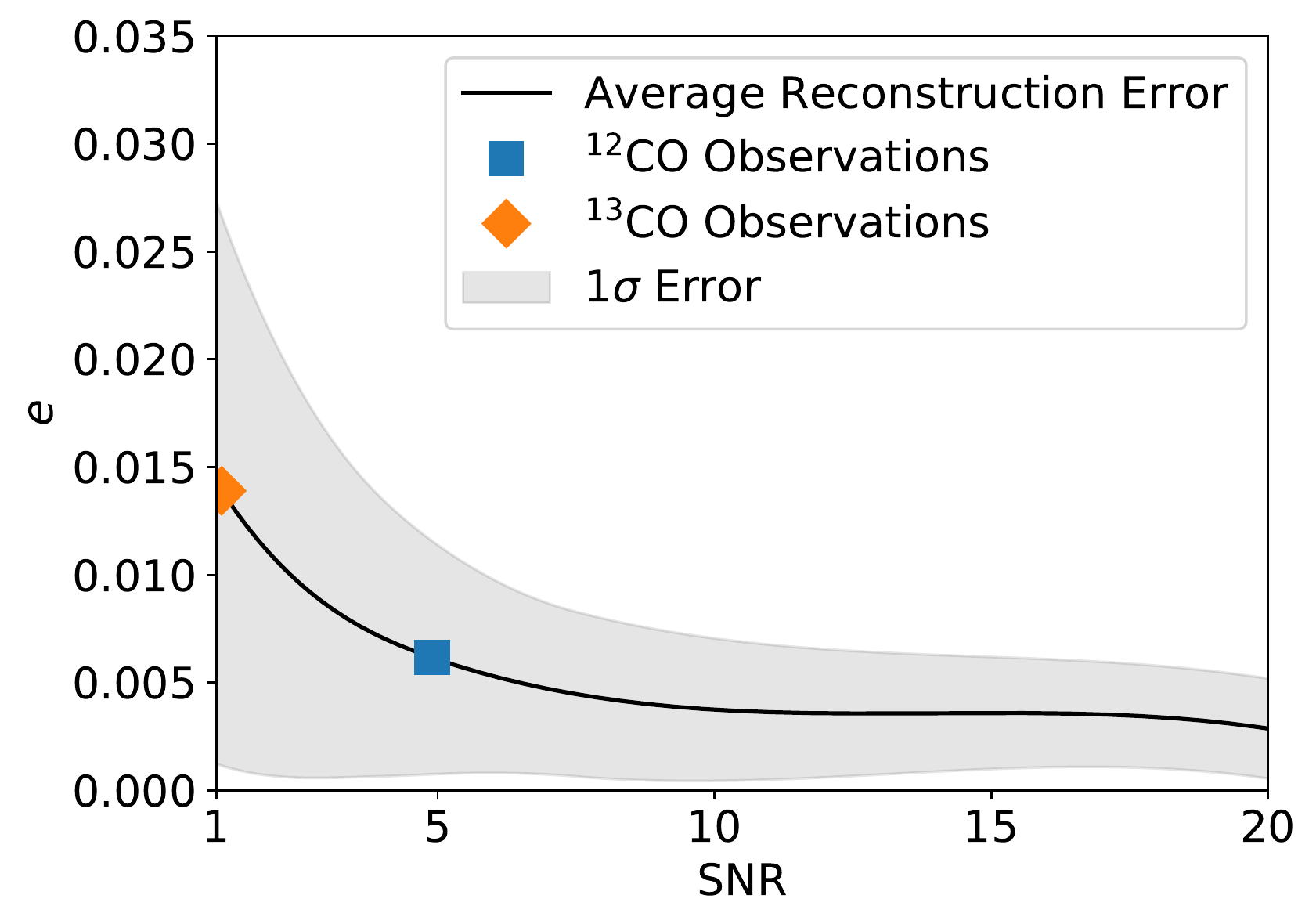}
        \caption{Dimensionless intensity reconstruction error for simulated images at each SNR. The solid line shows the median reconstruction error, and the shaded region shows the $1\sigma$ deviation for 1000 reconstructions at each SNR. The averaged measured SNR in each CO line observation is calculated and plotted. The blue square represents the median reconstruction error for ${}^{12}$CO observations. Similarly, the orange diamond represents the median reconstruction error for ${}^{13}$CO observations. The figure only shows 1$\leq$SNR$\leq$20 for the clarity of the behavior at low SNR. At higher SNR (SNR$>$20), we observe a further decreasing trend in $e$. It is clear that higher the SNR; better the reconstruction accuracy is achieved.}
        \label{fig:mock_reconstruction_error}
    \end{figure}

    The median intensity reconstruction error for each SNR of the 1000 simulated images is shown in Figure \ref{fig:mock_reconstruction_error}. The upper and the lower bounds of the error are one standard deviation from the median. We were able to achieve an average reconstructed intensity error within 0.01 (1\%) for images above SNR = 2.4 under the above explained set up. We expect the reconstruction accuracy to improve for higher SNR, as lower the noise, the higher the probability of convergence to the correct solution. The decreasing trend in reconstruction error is seen for increasing SNR. 
    
\subsection{Testing Reconstruction with Complete Images} \label{sec:error_free_reconstruction}
\begin{figure}
        \centering
        \includegraphics[width=\linewidth]{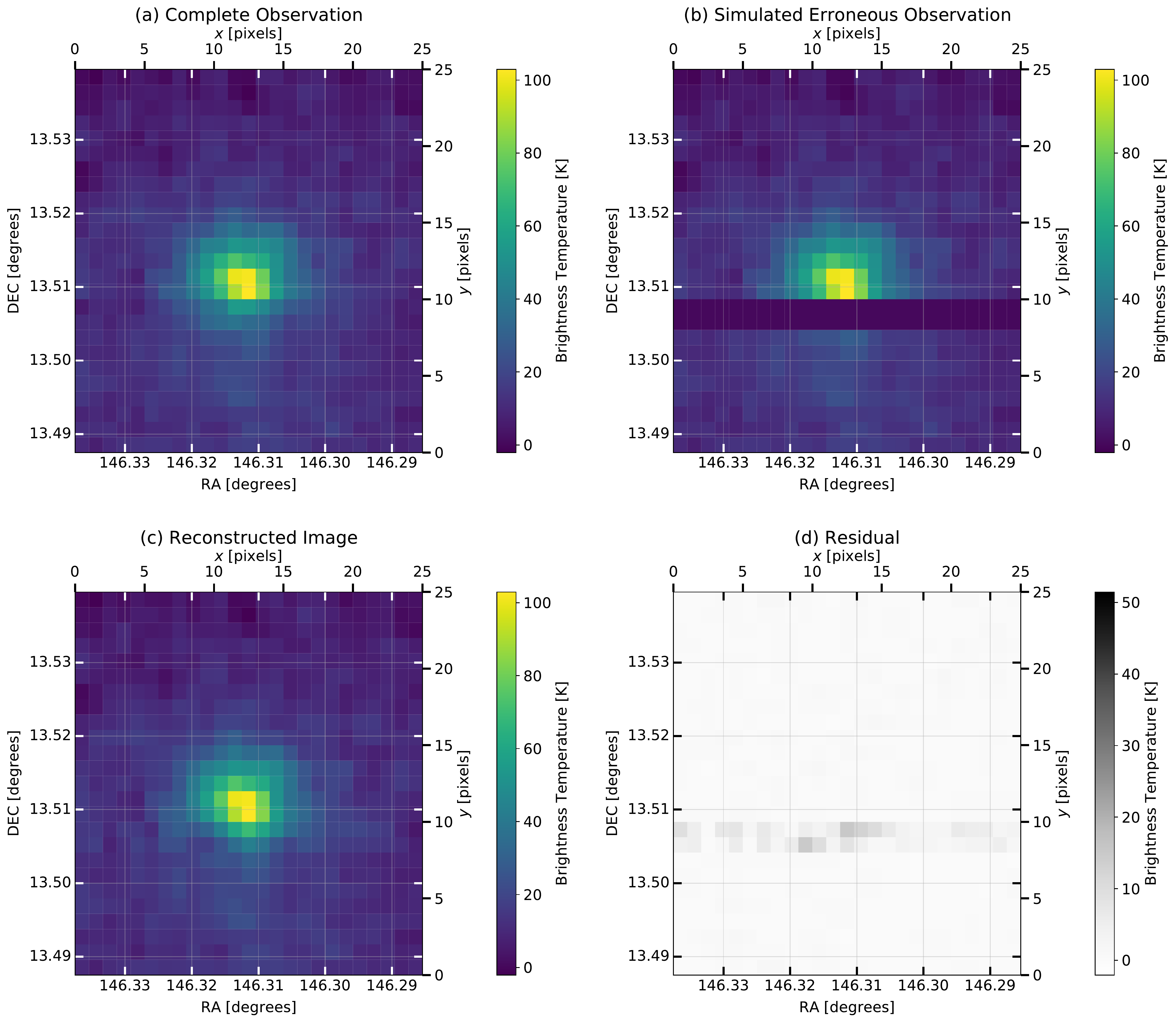}
        \caption{(a) is an complete observation done on 2018/04/19 18:49:01. (b) is the simulated erroneous observation where two commonly affected rows of pixels were masked. (c) is the reconstructed map created by running the reconstruction algorithm on (b). (d) is then the absolute residual between the reconstructed (panel (c)) and the original (panel (a)).}
        \label{fig:recon_accuracy}
    \end{figure}

    In the COMING project observations, not all calibration source images were affected by the detector error. In addition to testing the algorithm on noisy simulated data, we perform the same procedure on a set of complete calibration source observations, which we also used to generate mock images. We artificially mask the 9$^\textrm{th}$ and 10$^\textrm{th}$ pixel rows for each velocity integrated intensity map and restore them. Figure \ref{fig:recon_accuracy} is an example of the reconstruction of a complete calibration source image.
        
    Reconstruction errors for the 16 artificially masked and reconstructed images were analyzed according to Equation (\ref{eq:recon_error}). Table \ref{table:recon_accuracy} shows the results of the intensity reconstruction errors determined on the 16 artificially masked complete observations done on 2018/04/19 18:49:01. We were able to obtain an average error of 0.01061 and 0.01325 for \atom{\mbox{CO}}{}{12} and \atom{\mbox{CO}}{}{13} observations, respectively. These reconstruction errors are small in comparison to other uncertainties, such as atmospheric conditions. Therefore, in Section \ref{sec:reconstruction}, we employ the algorithm for the reconstruction of incomplete COMING images. 

    The SNR of the observations were measured to compare with the results from the simulated observations. \atom{\mbox{CO}}{}{12} and \atom{\mbox{CO}}{}{13} observations had an average SNR of 4.9 and 1.1, respectively. With this knowledge, we could estimate the errors in the measured intensities for our restored images. From the result in Figure \ref{fig:mock_reconstruction_error}, we expect the reconstructed intensity error to be 0.00616$\pm$0.00541 for \atom{\mbox{CO}}{}{12} and 0.01390$\pm$0.01276 for \atom{\mbox{CO}}{}{13}.

    \begin{table}
        \tbl{Reconstruction Errors for COMING images}{
        \centering
        \begin{tabulary}{\textwidth}{C C}
            \hline
            Image & Dimensionless Intensity Reconstruction Error ($e$)\\
            \hline \hline
            ${}^{12}$CO\_1\_1 & 0.00802 \\
            ${}^{12}$CO\_1\_2 & 0.01626 \\
            ${}^{12}$CO\_2\_1 & 0.01395 \\
            ${}^{12}$CO\_2\_2 & 0.02318 \\
            ${}^{12}$CO\_3\_1 & 0.00641 \\
            ${}^{12}$CO\_3\_2 & 0.00396 \\
            ${}^{12}$CO\_4\_1 & 0.00435 \\
            ${}^{12}$CO\_4\_2 & 0.00872 \\
            ${}^{13}$CO\_1\_1 & 0.00234 \\
            ${}^{13}$CO\_1\_2 & 0.01627 \\
            ${}^{13}$CO\_2\_1 & 0.00260 \\
            ${}^{13}$CO\_2\_2 & 0.01566 \\
            ${}^{13}$CO\_3\_1 & 0.00982 \\
            ${}^{13}$CO\_3\_2 & 0.04881 \\
            ${}^{13}$CO\_4\_1 & 0.00638 \\
            ${}^{13}$CO\_4\_2 & 0.00416 \\
            \hline
        \end{tabulary}}
        \label{table:recon_accuracy}
        \begin{tabnote}
            The comparison between the measured intensities in 11 $\times$ 11 pixel block of original and reconstructed images for the calibration source observation without the error. The average SNR for \atom{\mbox{CO}}{}{12} and \atom{\mbox{CO}}{}{13} images are 4.9 and 1.1, respectively. The average error measured for the images are 0.01061 and 0.01325 for \atom{\mbox{CO}}{}{12} and \atom{\mbox{CO}}{}{13} observations, respectively. Image names are formatted as "(CO line)\_(beam number)\_(polarization)". For example, an object observed in ${}^{12}$CO by beam 2 of polarization 1 is named "${}^{12}$CO\_2\_1".
        \end{tabnote}
    \end{table}

\subsection{Reconstructing Faulty Calibration Source Images} \label{sec:reconstruction}

    We apply the above-explained reconstruction method to the affected calibration sources in the COMING project. In total, there were six sets of 16 source images per night for two CO lines that the algorithm restored. In addition to the frequency shift artifacts, the detector errors in beams 2 and 3 resulted in up to 7 columns of pixels without information. The minimum number of pixels with information for reconstruction was 396 out of 625 ($25 \times 25$), which is about 63.3\% of the whole image. 
    
    We expect the Fourier nature of the observed calibration source not to change significantly daily. Thus, the parameters that were determined for the reconstruction of the complete observations (described in Section \ref{sec:error_free_reconstruction}) are used for the reconstruction of the distorted calibration source image.
    
    We examined whether the determined bandlimits satisfy the condition of $L \ge K$ (See Section \ref{sec:convergence}). When we assume a bandlimit of $U = U_{0.999, ^{12}{\rm CO}} = U_{0.985, ^{13}{\rm CO}}$ as explained in Appendix \ref{sec:appendix_bandlimit}, the number of Fourier components ($K$) to be estimated are 365. The least number of pixels with information ($L$) was 396. Thus, we confirm that we satisfy $L \ge K$ and that we can find a unique solution to the unmasking problem of COMING calibration source images.
    
    As mentioned in Section \ref{sec:error_free_reconstruction}, the averaged SNR for $^{12}$CO and $^{13}$CO were 4.9 and 1.1, respectively. By assuming that the faulty observations have similar SNR, the expected mean the intensity reconstruction errors are 0.00616 and 0.01390 with upper bounds of 0.01157 and 0.026650 for the reconstructed $^{12}$CO and $^{13}$CO observations in the COMING project, respectively. 
    
\begin{figure*}
        \centering
        \includegraphics[width=0.49\linewidth]{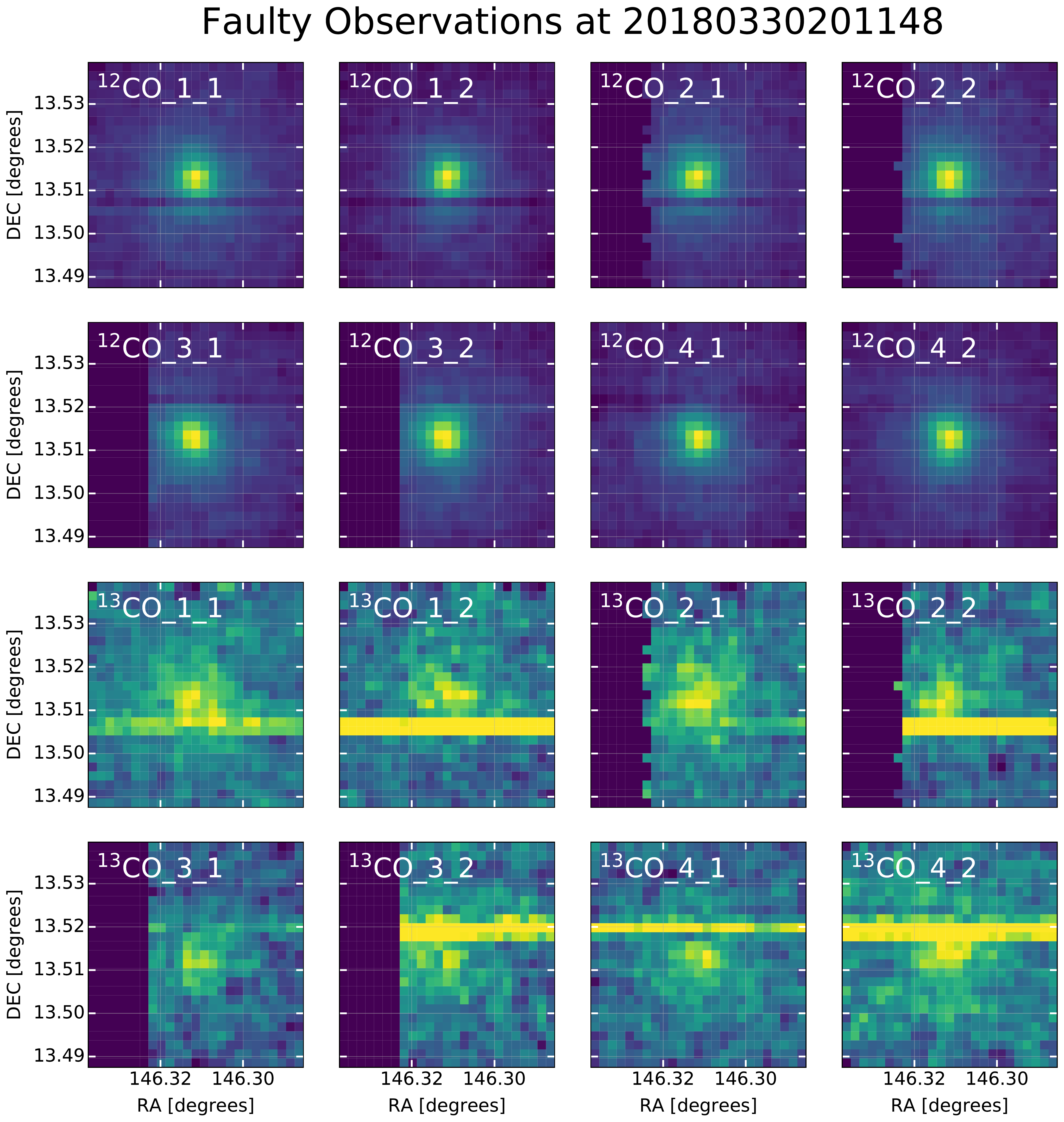}
        \hfill
        \includegraphics[width=0.49\linewidth]{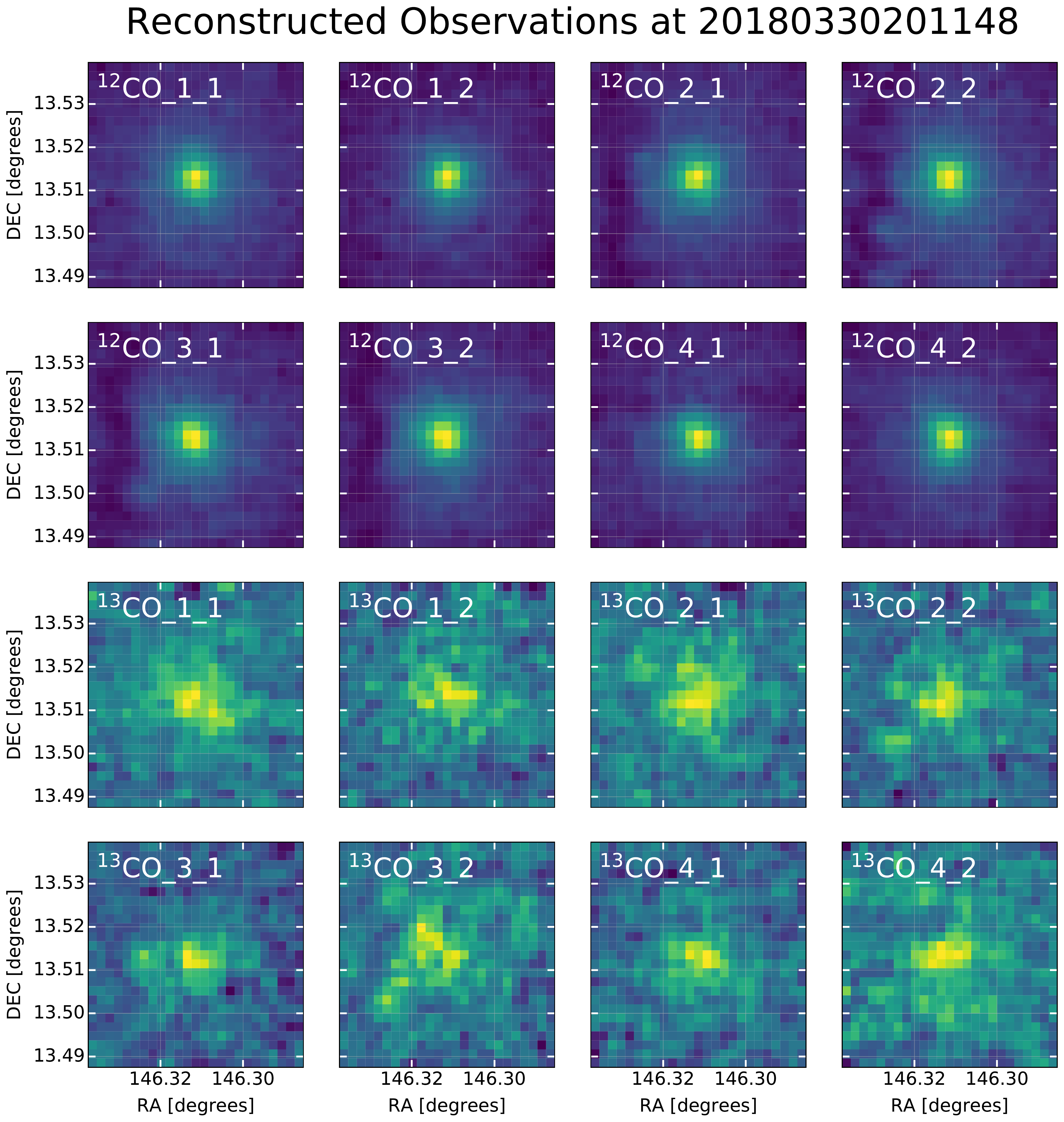}
        \caption{Left group of images show one set of observations (done on 2018/03/30 at 20:11:48) that was affected by the artifact. Right group of images are the corresponding reconstructed images of the ones on the left. Image names are formatted as "(CO line)\_(beam number)\_(polarization)". For example, an object observed in ${}^{12}$CO by beam 2 of polarization 1 is named "${}^{12}$CO\_2\_1". The color scales are not shown for clarity of the figure. However, the color scales of the original and its reconstructed for an image is the same.}
        \label{fig:orig_and_recon}
    \end{figure*}

    One set of observations (16 arrays) affected by the artifact is shown in Figure \ref{fig:orig_and_recon}. We stress that the reconstructed images show features that can never be recovered using any interpolation techniques. 

    The reconstructed images were used for the calibration of observations for 11 galaxies out of the total observed 147 galaxies. The COMING project overview paper \citep{Sorai2019} discusses how these calibration sources were used for the intensity calibrations of the galaxy maps in detail.

\section{Discussion on Possibilities and Limitations of the Reconstruction Algorithm} \label{sec:discussion}

    We restored the distorted intensity calibration source images in the COMING project successfully. However, it is crucial to understand the limitations and the possibilities of the reconstruction algorithm before the application to other astronomical images/signals. We should be particularly careful when deciding the bandlimits. 
        
    As explained in section \ref{sec:recon_algo}, for a signal to be bandlimited, it needs to contain information only of certain length scales and, therefore, finitely supported in Fourier space. We then have that the domain in the real space could not be finite. In this situation, a bandlimited signal extrapolation method such as the one introduced here can reconstruct incomplete astronomical signals. However, we need to be careful about bandlimitness for discrete signals. By the Nyquist-Shannon sampling theorem \citep{Nyquist_1928, Shannon1949}, a bandlimited continuous signal is representable without any error if the sampling rate is twice the highest frequency of the signal. In the case of images (which are discrete), each pixel corresponds to a sampling of a continuous signal. The relationship between the pixel size of the image and the maximum frequency of the continuous signal can then be represented as, $f_{\textrm{max}} = ({1}/{2\textrm{ pixels}})$. Similarly, if a discrete signal is to be bandlimited, the sample rate should be larger than the Nyquist rate. In other words, higher frequency components from the discrete Fourier transform should be zero ($F(u,v)=0$, where $\ M > |u| > U \ \& \  N > |v| > V$). In the case $F(u,v) \neq 0$ for $|u|=M$ or $|v|=N$, the smallest structures of the image will be the size of one pixel, and the signal will not be bandlimited. Additionally, we need to satisfy the condition $L \geq K$ as explained in Section \ref{sec:convergence}.

    A case where this reconstruction algorithm would fail is an image with just noise. For noise, each pixel is an independent realization of a random variable that follows a probability distribution. Such a signal contains information in all frequencies and will not be bandlimited. Therefore, introducing noise to a bandlimited signal will augment its characteristics. \cite{Sanz1983} have theoretically shown that even in the presence of noise, we can estimate a bandlimited signal with a controllable error. However, in real-world applications, it is essential to study the effects of noise in the reconstruction of the noisy astronomical images. 
    
    We analyzed the applicability of the technique under the above-discussed points. The Fourier nature of the signal was tested, as described in Appendix \ref{sec:appendix_bandlimit}. The reconstruction performance under noise was tested by generating mock images with varying noise levels, achieving reconstructed intensity error below 1\% for SNR $>$ 2.4. 
    
    We showed that for a high SNR bandlimited signal, the reconstruction algorithm is very capable of reconstructing partial signals. The reconstruction performance could be improved by incorporating noise information to the algorithm. Such modifications to the algorithm will be explored in future works.
    
    One of the strengths of the iterative procedure is the cheap computational cost, since estimating an inverse masking matrix can be computationally very expensive. Instead, the solution is estimated by successively operating the masking matrix. The computers are now powerful enough to invert significantly large matrices directly. However, this technique can be an alternative when the matrix is too large to be easily inverted. Additionally, the iterative error reducing nature also allows for easy implementation of additional constraints on the solution. Finally, the simplicity of the procedure allows the application in various contexts of astronomical signal reconstruction.

\section{Conclusion} \label{sec:conclusion}

    We have presented a mathematically consistent algorithm for unmasking astronomical signals. In the proposed algorithm, unmasking involves the inversion of the masking matrix under constraints on the Fourier components (bandlimited). The reconstruction algorithm bases on the core concepts of the Papoulis-Gerchberg algorithm (reconstruction algorithm for one-dimensional analytic functions), and extends to the case of astronomical images (two-dimensional discrete signals with noise). For intrinsically bandlimited signals, the presented algorithm can recover the complete information using only a partial number of pixels.
    
    We have demonstrated the application of the method to some of the intensity calibration source observations of the COMING project. Faulty $^{12}$CO and $^{13}$CO calibration source images had estimated reconstruction intensity errors of 0.616\% and 1.390\%, respectively, through repeated simulations. The restoration allowed for intensity calibration of CO multi-line maps for 11 galaxies out of 147 observed galaxies in the project. By this work, we have facilitated further scientific analysis of spatially resolved galaxy evolution studies. 
    
    We stress that the discussed extrapolation algorithm can restore structures smaller than the masked region. This ability is due to the estimation in Fourier space instead of real space and is fundamentally different from interpolation techniques. The stark contrast positions the reconstruction algorithm presented here as more promising and capable of the unmasking of signals than interpolation techniques.
    
    Unmasking images will be of great concern in the present, and upcoming large-scale imaging surveys like the LSST \citep{LSST_2019ApJ...873..111I} as subpar pixels are a common issue in CCDs. In the astronomical data-intensive age, the reconstruction algorithm discussed here positions itself as a high performing and computationally efficient algorithm to reconstruct missing regions of astronomical signals. 
    
    Reconstruction improvements are under study, and implementation of such modifications to the algorithm are possible. Such improvements will be reported in the future. Follow up applications and development of the technique will be of great importance to overcome many challenges in the upcoming astronomical studies.

\section{Funding}
    This work was supported in part by JSPS Grants-in-Aid for Scientific Research (17H01110 and 19H05076).
    This work was also supported in part by the Sumitomo Foundation Fiscal 2018 Grant for Basic Science Research Projects (180923), and the Collaboration Funding of the Institute of Statistical Mathematics ``New Development of the Studies on Galaxy Evolution with a Method of Data Science''.

\section*{Acknowledgements} 
\label{sec:acknowledgement}
    Firstly, we thank the referee, Shiro Ikeda, for valuable comments and constructive criticisms that significantly improved the quality of the article.  
    We thank the members of the COMING Project for providing the data and giving useful comments and suggestions in the analysis. 
    We would also like to thank Atsushi J.\ Nishizawa and Hiroyuki Tashiro for insightful discussions and comments on the algorithm and the manuscript. 

\section{References}
\bibliographystyle{bibstyle}
\renewcommand{\bibsection}{}
\bibliography{references}

\appendix
\section{Spectral Radius and the Expansiveness of an Operator} \label{sec:appendix_nonexpansive}

    Let there be an arbitrary matrix operator $\A:X \rightarrow X$ and $\BB:X \rightarrow X$ for $x \in X$. Their operator norm have the properties of,
    \begin{equation}
        \| \alpha \A \| = |\alpha | \cdot \| \A \| \textrm{, for } \alpha \textrm{ a scalar},
    \end{equation}
    \begin{equation}
        \| \A \cdot \BB \| \leq \| \A \| \cdot \| \BB \|.
    \end{equation}
    We now remind ourselves that the \textit{spectral radius} of a matrix $ \A $ is the largest absolute eigenvalue of $ \A $. 
    
    With the above, it is possible to associate the spectral radius and its operator norm. Let $\lambda$ be any eigenvalue of $ \A $ and $x \in X$ be any nonzero vector associated with the eigenvalue $\lambda$. Then we have that $ \A x=\lambda x$. By operator norm properties above, $\| \lambda x\|= |\lambda| \| x \|$ and $\| \A x \| \leq \| \A \| \cdot \| x \|$. Thus, for any eigenvalue $\lambda$ of matrix $ \A $,
    \begin{equation}
        \| \A \| \geq |\lambda |,
    \end{equation}
    which allows us to relate the spectral radius $\rho ( \A )$ as,
    \begin{equation}
        \| \A \| \geq \rho (\A).
    \end{equation}
    Then for a Hermitian matrix $\A$, $\| \A \| = \rho ( \A )$. The relation comes by definition of Hermitian ($ \A ^{\dagger} = \A $) and thus, $\| \A \|^2 = \rho ( \A ^{\dagger} \A ) = \rho ( \A ^2) = \rho^2 ( \A )$.
    
\section{Algorithmic Convergence for Discrete Signals} \label{sec:appendix_convergence}
    In this discussion, we consider the discrete case as we are interested in the reconstruction of images. For $\T$ defined in Eq. (\ref{eq:T_operator}), it is clear that $g$ does not affect the nonexpansiveness. Thus, we are concerned only about the $(\I - \M_{\Gamma})\B$ term. Let us define $\R = (\I - \M_{\Gamma}) \B$. We can write the Euclidean norms for iterations $i$ and $j$ as, 
    \begin{eqnarray}
        \left\| \R g_{i} \right. &-& \left. \R g_{j}\right\| = \left\|(\I - \M_{\Gamma}) \left(\B g_{i}- \B g_{j} \right)\right\|  \nonumber \\
        &=& \left\{\sum_{x,y}(1-m(x, y)) \left(\tilde{g}_{i}(x, y)-\tilde{g}_{j}(x, y)\right)^{2}\right\}^{1 / 2} ,
    \end{eqnarray}
    where,
    \begin{equation}
        m (x, y) = \left \{ \begin{array}{ll}
                1  &  \  \textrm{if } (x, y) \in \Gamma \\
                0   &  \ \textrm{elsewhere} ,
        \end{array} \right. 
    \end{equation}
    and $\tilde{g}_{i} = \B {g}_{i}$ and $\tilde{g}_{j} = \B {g}_{j}$. As R is a linear operator, the summation in $(x, y)$ coordinates can be separated as follows,
    \begin{eqnarray}
    \left\|\R g_{i} - \R g_{j} \right\| &=& \left\{\sum_{x, y} \left(\tilde{g}_{i}(x,y)-\tilde{g}_{j}(x,y)\right)^{2}\right. \nonumber \\ &-& \left. \sum_{(x,y) \in \Gamma }\left(\tilde{g}_{i}(x,y) - \tilde{g}_{j}(x,y)\right)^{2}\right\}^{1 / 2} .
    \end{eqnarray}
    From above we can straightforwardly write,
    \begin{equation}
        \left\|\R g_{i} - \R g_{j} \right\| \leq \gamma_1 \left\|\B g_{i} - \B g_{j} \right\| ,
    \end{equation}
    where $0 \leq \gamma_1 \leq 1$ with $\gamma_1=1$ only when $g_{i} = g_{j}$ in the observed region $\Gamma$ (second summation becomes zero). Then by the Parseval's theorem for DFT, we get the following;
    \begin{eqnarray}
        \left\|\B g_{i} \right. &-& \left. \B g_{j} \right\| = 
        \left\{ \frac{1}{NM} \sum_{u,v} 
        \left( \beta_\Omega G_i(u,v) - 
        \beta_\Omega G_j(u,v) \right)^2 \right\}^{1 / 2} 
        \nonumber \\
        &=& \left\{ \frac{1}{NM} \sum_{(u,v) \in \Omega} 
        \left( G_i(u,v) - G_j(u,v) \right)^2
        \right\}^{1 / 2} ,
    \end{eqnarray}
    where {$G_i(u,v) = \F [g_i(x, y)]$ and $G_j(u,v) = \F [g_j(x, y)]$}. Clearly we can then write,
    \begin{equation}
        \left\|\B g_{i} - \B g_{j} \right\| \leq \gamma_2 \left\| g_{i} -  g_{j} \right\| ,
    \end{equation}
    where $0 \leq \gamma_2 \leq 1$ with strict equality only when $G_i(u,v) = G_j(u,v)$ outside $\Omega$.
    We now have that both $(\I-\T)$ and $\B$ are both nonexpansive. We can combine the two relations to write,
    \begin{equation}
        \left\|\R g_{i} - \R g_{j} \right\| \leq \gamma_1\gamma_2 \left\| g_{i} -  g_{j} \right\| ,
    \end{equation}
    where $0 \leq \gamma_1 \gamma_2 \leq 1$. Thus, implying from the definition of $\R$ that $(\I - \M_{\Gamma}) \B$ is also nonexpansive. 
    
    In the case of continuous signals, integrals replace the summations in the above norm manipulations. For such a case, we could show that $\gamma_1$ and $\gamma_2$ could not simultaneously be unity. We then have that $0 \leq \gamma_1 \gamma_2 < 1$, and the $(\I - \M_{\Gamma}) \B$ operator strictly nonexpansive as shown by \citet{Landau_1961}. Such strictly nonexpansiveness guarantees the convergence of the algorithm to a unique fixed point \citep{Ortega2000}. \cite{Schafer1981} discusses the difficulty for strict nonexpansiveness of $(\I-\M_{\Gamma})\B$ in the discrete case results in the algorithm be more sensitive to noise and not converge. However, we could note that $\gamma_1 \gamma_2$ is identity only when $\B {g}_{i} = \B {g}_{j}$ for some iteration $i$ and $j$. In such a situation, the difference ($\R g_{i} - \R g_{j}$) would also be bandlimited and identically zero. Such a condition violates the Paley-Weiner theorem, which states that the Fourier transform of the signal should be compactly supported (bandlimited) for a square-integrable function to be well-defined \citep{Wiener_1934, Rudin1987}. By showing $(\I - \M_{\Gamma}) \B$ is strictly nonexpansive, we have shown that $\T$ also is strictly nonexpansive. A strictly nonexpansive operator exhibits a convergence to a unique solution.

\section{Determination of Constraints for Reconstruction} \label{sec:appendix_bandlimit}
    
    \begin{figure*}
        \centering
        \includegraphics[width=\linewidth]{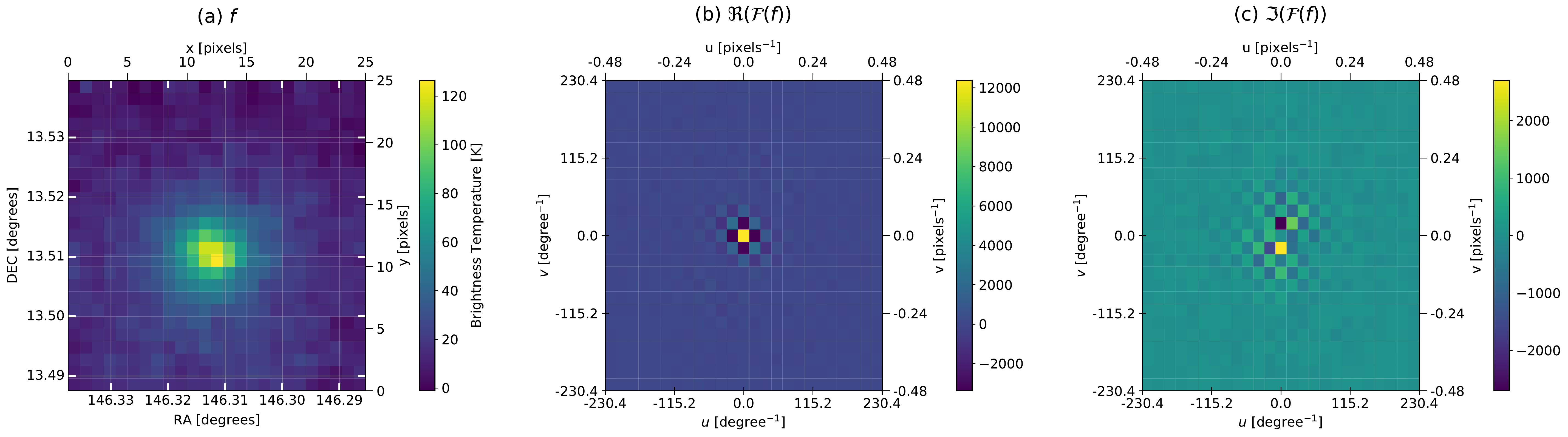}
        \caption{{(a) is an complete observation done on 2018/04/19 18:49:01 (b) and (c) are the real and the imaginary components of the Fourier transform of (a) respectively. It is seen that the signal observed is dominated by low-frequency components.}}
        \label{fig:fourier_signal}
    \end{figure*}
    
    \begin{figure}
        \centering
        \includegraphics[width=0.95\linewidth]{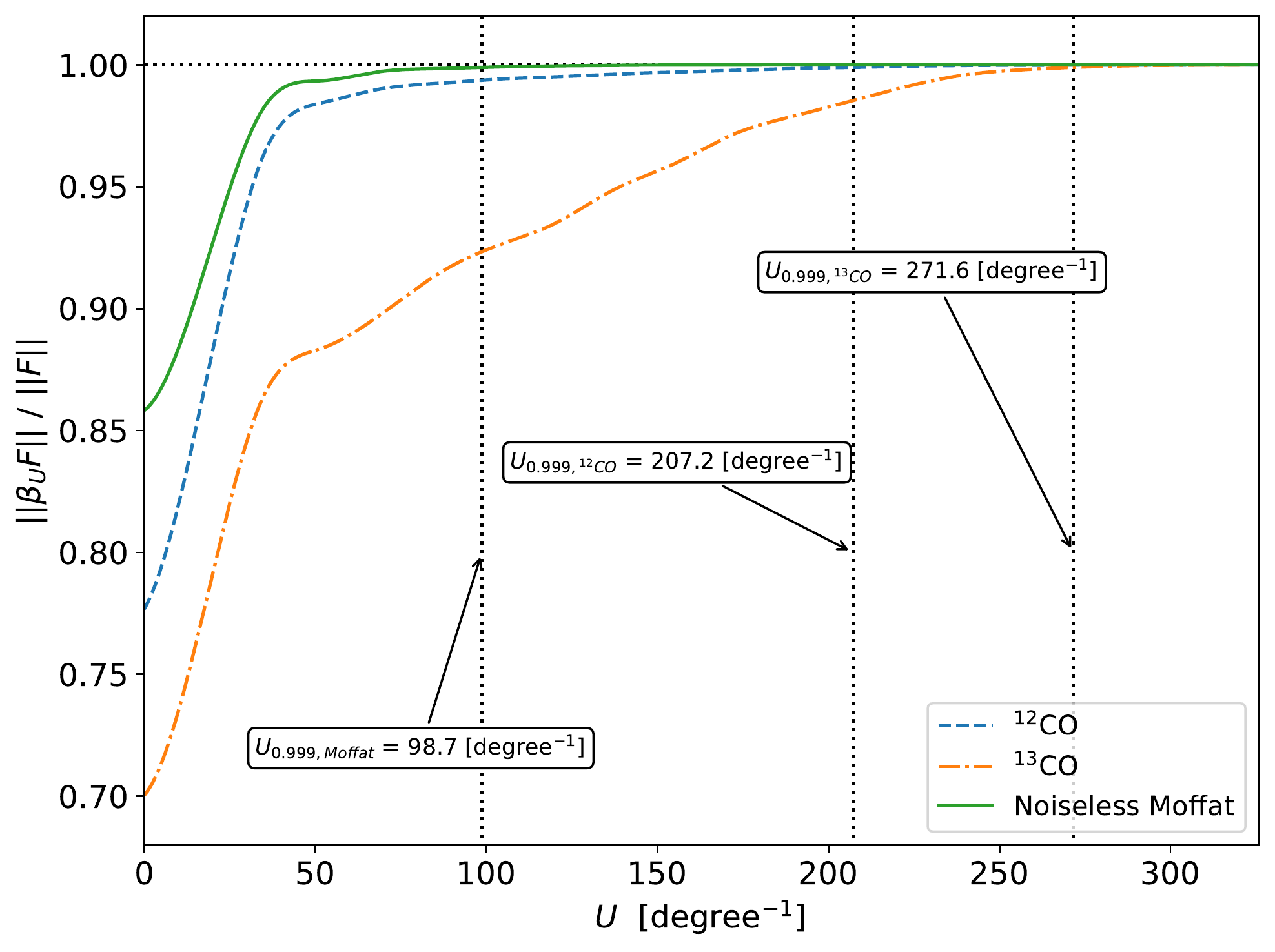}
        \caption{{The figure shows the fraction of the band-limited Fourier image $\ell_2$-norm enclosed in radius $U$. The solid line shows the case for the noiseless Moffat model image. Dashed and dash dotted lines corresponds to the average Fourier profiles of $^{12}$CO and $^{13}$CO observations. The vertical dashed lines correspond to the values of $U$ where 99.9\% of the $\ell_2$-norm is enclosed for each kind of observation.}}
        \label{fig:bandlim}
    \end{figure}
    
    The inversion of the mask requires known information about the underlying signal. The bandlimited assumption regularizes the iterative inversion in our algorithm. For idealized signals, we can theoretically determine the Fourier support. However, we need to evaluate suitable bandlimits for real-world signals (e.g., noisy) that we reconstruct. This section discusses a procedure for the determination of the constraints for reconstruction.

    Analyzing the distribution of the signal in the Fourier space gives us the bandlimits. We calculate the fraction of the signal included in the defined bandlimits. Figure \ref{fig:fourier_signal} shows an example of a COMING calibration source image and its real and imaginary components in the Fourier space. Analyzing complete images do not include the masking effect in the transform space. The reconstruction of incomplete images uses the determined bandlimits, assuming that the complete and incomplete signals have the same characteristics. 
    
    We define the bandlimiting operator $\beta_\Omega$ with the $\Omega$ defined as, 
    \begin{equation}
        \Omega = \{ (u, v) \ | \ ( u^2 + v^2 \leq U ) \} .
    \end{equation}
    The bandlimiting operator defined above is a low-pass filter. We applied the above condition with a function of radius because the images were $N \times N$, where $N=25$. We denote the bandlimiting operator $\beta_\Omega$ for the above $\Omega$ as $\beta_U$ to simplify the notation.
    
    We calculated the fraction of the $\ell_2$-norm of the signal within a frequency $U$ with respect to the total $\ell_2$-norm of the signal in the transform space. Mathematically we can write the above as, ${||\beta_U F||}\ /\ {||F||}$. For an ideal bandlimited signal such as the 2D sinc function, we can define $U$ where ${||\beta_U F||}\ /\ {||F||} = 1$. However, noise extends the bandlimits by introducing components of other frequencies. Thus, we defined the bandlimit constraints for reconstruction as $U$ where ${||\beta_U F||}\ /\ {||F||} < 1$. In this analysis, we determined $U$ where 99.9\% of the $\ell_2$-norm in Fourier space is enclosed (i.e. ${||\beta_U F||}\ /\ {||F||} < 0.999$) for each of the signal.
    
    The above definition for bandlimit will include the frequencies from noise. The reason to include as many frequency components is that we lose the total intensity of the signal by filtering out frequencies. Applying a bandlimit operator that filters out non-signal frequencies results in lower intensities for our reconstructed intensity calibration source images. 
    
    We calculate the bandlimiting operator $\beta_U$ for $U=0$ to $U=\sqrt{2}U_{\textrm{max}}=325.8\;[\textrm{deg}^{-1}]$ where $U_{\textrm{max}} = 230.4\;[\textrm{deg}^{-1}]$ (Nyquist-Shannon sampling theorem). For frequencies $U \geq U_{\textrm{max}}$, we begin to consider scales smaller than the pixel size. Reconstruction algorithms will not converge by employing such constraints as masking boundaries are also of a single pixel scale. Then the algorithm is unable to get rid of the masking effect at each estimation. We should therefore set the bandlimit constraint for reconstruction as $U \leq U_{\textrm{max}}$ as discussed in Section \ref{sec:discussion}.
    
    The bandlimit estimation was done for the noiseless Moffat model, and the 16 error-free calibration source images ($^{12}$CO and $^{13}$CO) observed on 2018/04/19 18:49:01. These values were denoted as $U=U_{0.999, {\rm Moffat}}$, $U=U_{0.999, ^{12}{\rm CO}}$ and $U=U_{0.999, ^{13}{\rm CO}}$ for the noiseless Moffat model, $^{12}$CO and $^{13}$CO observations respectively. The 16 estimated bandlimits corresponding for each array in the detector channels were used for reconstruction of the distorted images from the same channel.
    
    In the analysis, each kind of observations ($^{12}$CO and $^{13}$CO) had similar profiles in Fourier space. Therefore, we considered the average of the eight arrays in each CO observation. Figure \ref{fig:bandlim} shows the result of the $\ell_2$-norm distribution for radius U from the origin in Fourier space for the COMING images.
    
    The bandlimit $U_{0.999, ^{13}CO} = 266.4 [\textrm{deg}^{-1}]$ could not be used for the reconstruction because $U_{0.999, ^{13}{\rm CO}} \geq f_{\textrm{max}}$. As mentioned earlier, for frequencies $U \geq f_{\textrm{max}}$, scales smaller than the pixel size will be included. Therefore a different bandlimit constraint had to be decided to reconstruct $^{13}$CO observations. We employed the same constraint ($U_{0.999, ^{12}CO}$) for $^{13}$CO observations. We justify the use of this value for the reconstruction of $^{13}$CO observations because ${||\beta_{U} F||}\ /\ {||F||} = 0.985$ for $U=U_{0.999, ^{12}{\rm CO}}$.

\end{document}